\documentclass[useAMS,singlespacing]{mnras}
\usepackage{graphicx,amssymb}

\usepackage{newtxtext,newtxmath}
\usepackage{xcolor}
\usepackage{ulem}

\title[Cosmic ray acceleration and transport with magnetic mirroring]
{Cosmic ray transport and acceleration with magnetic mirroring}

\author[AR Bell, JH Matthews, AM Taylor, G Giacinti]
{AR Bell$^{1,2}$\thanks{E-mail:Tony.Bell@physics.ox.ac.uk},
JH Matthews$^{3}$,
AM Taylor$^{4}$,
G Giacinti$^{5,6}$
\\
$^{1}$University of Oxford, Clarendon Laboratory, Parks Road, 
Oxford OX1 3PU, UK\\
$^{2}$Central Laser Facility, STFC Rutherford Appleton Laboratory, Oxfordshire OX11 OQX, UK\\
$^{3}$Department of Physics, Astrophysics, University of Oxford, Denys Wilkinson Building, Keble Road, Oxford, OX1 3RH, UK\\
$^{4}$Deutsches Elektronen-Synchrotron, Platanenallee 6, 15738 Zeuthen, Germany\\
$^{5}$Tsung-Dao Lee Institute, Shanghai Jiao Tong University, Shanghai 201210, People's Republic of China\\
$^{6}$School of Physics and Astronomy, Shanghai Jiao Tong University, Shanghai 200240, People's Republic of China\\
}
\begin{document}
\date{\today}
\pagerange{\pageref{firstpage}--\pageref{lastpage}} \pubyear{2023} 
\maketitle
\label{firstpage}
\begin{abstract}
We analyse the transport of cosmic rays (CR) in magnetic fields that are structured on scales greater than the CR Larmor radius.
We solve the Vlasov-Fokker-Planck (VFP) equation for various mixes of mirroring and small-angle scattering and show that relatively small deviations from a uniform magnetic field can induce mirroring and inhibit CR transport to levels that mimic Bohm diffusion in which the 
CR mean free path is comparable with the CR Larmor radius.
Our calculations suggest that shocks
may accelerate CR to the Hillas (1984) energy without the need for magnetic field amplification on the Larmor scale.
This re-opens the possibility, subject to more comprehensive simulations, that
young supernova remnants may be accelerating CR to PeV energies,
and maybe even to higher energies beyond the knee in the energy spectrum. 
We limit our discussion of CR acceleration to shocks that are non-relativistic.
\end{abstract}
\section{ Introduction}
There has been considerable progress in understanding the origins of cosmic rays (CR) during the nearly half century
since the development of the theory of diffusive shock acceleration (DSA) 
by Krymsky (1977), Axford et al (1977), Bell (1978) and Blandford \& Ostriker (1978).
Observations across the whole range of energies have improved immensely and we now have a credible theoretical understanding of how and where CR are accelerated in the range of energies up to 100TeV.
However, there are many questions to be answered about the origins of Galactic CR at energies of PeV and above.
At the very highest energies, 1-200EeV, CR must originate outside the Milky Way Galaxy since these have a Larmor radius exceeding the size of the Galaxy.
Observations of ultra-high-energy CR (UHECR) are limited by their rarity and the lack of secondary radiation, 
although the Pierre Auger Observatory (PAO) and the smaller Telescope Array (TA) are beginning to provide 
composition and anisotropy data that constrain their origin (Tsunesada et al 2021, Plotko et al 2023).
The most challenging remaining mystery is the origin of Galactic CR at and above the knee in the energy spectrum at a few PeV.

From both observation and theory, it appears that CR acceleration by the historical supernova remnants (SNR)
tails off at a few hundred TeV (Zirakashvilii \& Ptuskin 2008, Bell et al 2013, Cristofari 2021).
One possibility is that PeV CR are accelerated by SNR in their first decades when a high velocity shock expands into a dense circumstellar medium.
Other possibilities are that PeV accelerated near the Galactic centre are transported to the Earth without being lost from the Galaxy (Abramowski et al 2016, Adams et al 2021, Muena et al 2024),
or that star-forming regions with strong stellar winds or repeated SNR expanding into a strong magnetic field provide enhanced conditions 
for first or second order Fermi acceleration beyond 1 PeV (Vieu et al 2022, Vieu \& Reville 2023, Vink 2024, Aharonian et al 2024).

We use S.I. units throughout this paper. except that eV is used for CR energy and values of the magnetic field are quoted in Gauss.
The Hillas energy $E=ZuBR$ (Lagage and Cesarsky 1983a,b, Hillas 1984)  imposes a basic limit on the maximum energy to which CR can be accelerated.
In this expression, $Ze$ is the particle charge, $B$ is the magnetic field, $R$ is the size of the accelerating region, and $E$ is the CR
energy.
A more incontrovertible but less stringent limit that  $E<ZcBR$ arises from the need for the Larmor radius to be less than $R$.
The tighter limit $E<ZuBR$ arises from the need for the distance over which acceleration occurs to be smaller  than $R$ or for the time required for acceleration to be less than the available time $R/u$.
One form of the spatial constraint is that a CR passing a distance $R$ through the MHD electric field $-{\bf u}\times {\bf B}$ 
traverses a maximum potential difference $uBR$.
This limit most clearly applies to acceleration at perpendicular shocks where the magnetic field
is perpendicular to the shock normal.
It also applies to neutron star magnetospheres and nebulae where $uBR$ is the potential difference between pole and equator.
In principle, the path of a CR through turbulence may sample the MHD electric field preferentially to gain an energy 
exceeding $ZuBR$, but this is difficult to arrange.  
To reach the higher Larmor limit $E=ZcBR$, a CR trajectory would need to be 
microscopically aligned with the $-{\bf u}\times {\bf B}$   MHD electric field throughout a time $R/u$
during which the CR travels a distance $(c/u)R$.
The Hillas limit, $E<ZuBR$, is most easily understood as a competition between the acceleration rate and the available timescale
$t_{\rm limit}=R/u$ as discussed below.  
A recent discussion of the general applicability of the Hillas condition is given by Oka et al (2024).

The maximum energy to which diffusive shock acceleration (DSA) 
can accelerate CR is determined by CR transport, especially in the upstream plasma.
Fluctuations in the magnetic field scatter the CR  and stop them escaping far upstream of the shock.
Scattering is strongest when the magnetic field is structured on the scale of the Larmor radius $r_g$ and scatters CR with a mean free path 
on the scale of  the Larmor radius.
Scattering on the Larmor scale leads to Bohm diffusion with a diffusion coefficient $D_B\sim r_gc$.
In DSA at non-perpendicular shocks, a condition for reaching the Hillas energy is that CR diffuse with the Bohm diffusion coefficient.

As shown by Lagage \& Cesarsky (1983a,b), even with Bohm diffusion, 
the characteristic ambient magnetic field in the interstellar medium (ISM) of 1-30$\mu$G
is insufficient for acceleration to 1PeV.
Magnetic field amplification by a non-resonant instability (Bell 2004) can provide the hundred-fold increase 
needed for the Hillas limit to reach 1PeV.
Given sufficient time to evolve, non-resonant amplification saturates on the spatial scale of the Larmor radius of the CR driving the instability
as required.
However, the growth rate is too slow for the instability to grow to the Larmor scale of CR with PeV energies
(Zirakashvilii \& Ptuskin 2008, Bell et al 2013).

Standard DSA allied with non-resonant magnetic field amplification may possibly account for PeV 
acceleration in special environments such as SNR in their first few decades of existence, the Galactic centre, or young stellar clusters,
but it does not provide a basis for CR acceleration to 1PeV in any currently observed SNR in the Galaxy.
 
Here we examine the possibility that CR transport might be inhibited by magnetic mirrors which 
confine CR more closely to the shock in the upstream plasma and thereby facilitate more rapid acceleration.
This mirror-dominated shock acceleration (MDSA) may ease the challenge of explaining Galactic CR acceleration to PeV,
and may even open up the possibility of CR acceleration beyond the Hillas limit.

Many authors have recognised that simple diffusion theory, as assumed in derivations of the Hillas limit, 
does not tell the whole story since CR trajectories follow magnetic field
lines if their Larmor radius is smaller than distances over which the magnetic field changes in magnitude or direction.
This can lead to  sub-diffusion (Duffy et al 1995), super-diffusion (Lazarian \& Yan 2014),   cross-field diffusion in perpendicular shocks (Jokipii 1982, 1987), and trapping in magnetic loops (Decker 1993).
The contribution of magnetic mirrors to CR transport has recently received growing attention
(Lazarian \& Xu  2021,
Barreto-Mota et al 2024,
Zhang \& Xu 2024,
Reichherzer et al 2025)

The possibility of  CR acceleration when trapped between an upstream mirror and a shock, or between an upstream mirror and  a downstream mirror,
was considered by Jokipii (1966) in the context of the solar wind.
Jokipii showed that rapid acceleration is possible leading to CR energy increases that are essentially adiabatic
as the distance contracts between an upstream  mirror and the shock or a mirror downstream of the shock.

The problem with mirrors for general acceleration is that CR are too rapidly swept away downstream by the mirror, 
and that mirroring only applies to CR with particular pitch angles, 
thus limiting the overall acceleration efficiency and failing to produce the
extended $p^{-2.3-2.6}$ spectrum required to explain Galactic CR.
Jokipii's theory was developed before the advent of modern DSA theory which showed that CR return to the shock 
typically $ c/u$ times, and some CR return to the shock many more times as determined by diffusion theory or random walk theory.
If mirror reflection confines some CR more closely to the shock while still
allowing CR to the traverse the shock many times in an overall diffusive manner, 
then energies of a few CR may be boosted to, and possibly beyond, the Hillas limit by rapid reflection.
CR acceleration by MDSA  may be less efficient, and this would be consistent with the observed steepening at the knee.

Jokipii (1966) noted that non-conservation of the magnetic moment at a shock, might lead to de-trapping of CR, allow
CR to stay with the shock for longer and be accelerated to higher CR energies.
It is a variant on this line of thinking that we pursue in this paper.
Angular scattering by sub-Larmor-scale fluctuations in the magnetic field might allow 
CR to pass semi-randomly through mirrors.
CR  might benefit from multiple acceleration episodes as a population of mirror-trapped CR
is overtaken by a shock advancing into a plasma containing mirrors.
In section 5 to 8 below we examine transport when there is both mirroring and small-scale scattering.

\section{Cosmic ray transport through a single mirror: the equations}

We begin with two idealised analyses (sections 2 to 4) to set the scene for the more comprehensive analysis that follows, 
and to offer physical explanations to underpin our computational model.
First, we analyse CR interaction with a mirror when two conditions both apply: (i) no small-angle scattering 
and (ii) the mirror scalelength exceeds the Larmor radius, $L\gg r_g$, such that cross-field drifts are small
and the magnetic moment is conserved.  
We consider only highly relativistic protons ($Z=1$).  
The theory can easily be
extended to nuclei with $Z>1$.  
CR travel along magnetic field lines at a speed $\mu {\rm v}$, where $|{\bf v}|=c$, and circulate around the field line 
at speed ${\rm v}_\perp = (1- \mu^2)^{1/2}{\rm v}$ where $\mu=\cos \theta$ and $\theta$ is the
angle between the CR velocity and the local magnetic field.
The gyration radius is $({\rm v}_\perp /c) r_g$ where $r_g = p/eB$.
In this section we use $x$ as the local direction parallel to $\bf B$. 
The instantaneous separation (in $y$ and $z$) of the CR from its gyrocentre on its magnetic field line we call ${\bf r}$.

${\bf B}_\parallel $ ($B_\parallel =|{\bf B}|$) is the magnetic field on the field line around which the CR circulates.
We assume in this section that the magnitude of ${\bf B}_\parallel (x)$ varies along the magnetic field line but its direction does not change.
Variation in direction can be added to the theory, but it adds only small terms if $\bf B$ varies 
over distances much greater than $r_g$.

In order to maintain $\nabla. {\bf B}=0$, we must include the variation of $\bf B$ (with
components $B_y$ and $B_z$)
in the perpendicular $y$ and $z$ directions.
We follow a conventional analysis of CR trajectories along field lines and through a mirror.
The instantaneous magnetic field experienced by a CR on sub-Larmor timescales is
$$
{\bf B}'= {\bf B}_\parallel+ {\bf r}.\nabla {\bf B}
\hskip 0.3 cm {\rm where} \hskip 0.3cm
{\bf r}=- \frac {p}{ce B_\parallel^2}  {\bf v}_\perp \times {\bf B}_\parallel
\eqno(1)
$$
The CR experiences a force ${\bf F}_\parallel=e{\bf v}_\perp \times {\bf B}_\perp '$
in the $x$ direction where
${\bf B}_\perp '= {\bf r}.\nabla {\bf B} _\perp$ is the instantaneous vector component perpendicular to the field line. 
$$
{\bf F}_\parallel=
- \frac {p}{cB_\parallel ^2} {\bf v}_\perp \times 
\big [ ( {\bf v}_\perp \times {\bf B} ).
 \nabla \big ]
{\bf B}_\perp
\eqno(2)
$$
When averaged over the gyration around the magnetic field line, this reduces to
$$
F_x=
\frac {p {\rm v}_\perp^2}{2c B_\parallel}
\left ( \frac {\partial {B_y}}{\partial y}+\frac {\partial {B_z}}{\partial z}\right )
\eqno(3)
$$
Applying $\nabla . {\bf B}=0$ gives
$$
F_x=-
\frac {pc (1- \mu ^2)}{2 B_\parallel } \frac {\partial B_\parallel}{\partial x}
\eqno(4)
$$
In the absence of electric field, the Vlasov equation for the CR distribution function takes the form
$$
\frac {\partial f}{\partial t}  + c\mu \frac {\partial f}{\partial x} 
+F_x \frac {\partial f}{\partial p_x} 
=0
\eqno(5)
$$
With minor algebraic manipulation, this becomes
$$
\frac {\partial f}{\partial t}  + c\mu \frac {\partial f}{\partial x} 
-  \frac {c}{2 B} \frac {\partial B}{\partial x} (1- \mu ^2) \frac {\partial f}{\partial \mu} 
=0
\eqno(6)
$$
Equation (6) is a well-known gyrokinetic equation neglecting cross-field drifts and using the notation that 
$\mu$ is the cosine of the angle between the CR velocity and the $x$ direction.
$\mu$ is not the magnetic moment (a notation used in other analyses of mirroring).
We represent the normalised magnetic moment for a monoenergetic CR distribution by the symbol $M$:
$$
M=(1-\mu^2)/B
\eqno(7)
$$
It can easily be demonstrated by substitution in equation (6) that $f=f(M)$ is a steady state solution 
as expected for consistency with conservation of the magnetic moment.

Also, it can easily be shown by integrating equation (6) that the CR current density is 
proportional to $B$ in steady state.
This dependence on $B(x)$ is correct because the cross-sectional area associated with a magnetic field line is proportional to $1/B(x)$.  The current density in the mirror has to increase in
proportion to $B(x)$ to carry the same total current.

\section{Transport through a single magnetic mirror}

We now apply magnetic moment conservation, $f=f(M)$, to a simple idealised case.
We solve for $f$ along a magnetic field line between two points, $x=-L$ and $x=L$, at which $B=B_{0}$.
We refer to these points as `isotropic plates' for reasons now to be explained.
The magnetic field is larger between the plates ($-L<x<L$) with a maximum  $B_{\max}$ at the midway point $x=0$.
We assume that there is no CR angular scattering in the space $-L<x<L$.
We 
assume that CR injected into the space $-L<x<L$ from either of the plates are part of a locally isotropic monoenergetic distribution.
The difference between the plates is that the CR density is larger at the left-hand plate than at the right-hand plate.
The magnetic field is spatially symmetric about $x=0$.

From the assumption of isotropy, $f$ is the same for all CR emerging from the left-hand plate.
This allows us to set 
$f=f_L$ for $\mu>0$ at $x=-L$.
Similarly at the right-hand plate, $f=f_R$ for $\mu <0$.
Some of the CR injected from the left-hand plate, with $\mu$ less than some value $\mu_L$ (see figure 1), are reflected by the magnetic mirror and return to the left-hand plate with $\mu$ reversed in sign, but with the same $\mu ^2$ to conserve the magnetic moment $M$, and therefore having the same $f$.
Hence, at the left-hand plate,  $f=f_L$ for all $-\mu _L< \mu < 1$.

CR arriving at the left-hand plate with $\mu<- \mu _L$ have been transmitted through the mirror from the right-hand plate, and therefore have $f=f_R$ where $f_R$ (smaller than $f_L$) is the value of $f$ for CR emitted from the right-hand plate.
Consequently, at the left-hand plate,
$$
f(\mu)=f_L \hskip 0.1 cm {\rm for } \hskip 0.1 cm - \mu_L< \mu <1
\hskip 0.2 cm ; \hskip 0.2 cm 
f(\mu)=f_R \hskip 0.1 cm {\rm for } \hskip 0.1 cm -1< \mu < -\mu_L
\hskip 5 cm 
\eqno(8)
$$
Similarly at the right-hand plate 
$$
f(\mu)=f_R \hskip 0.1 cm {\rm for } \hskip 0.1 cm -1<\mu <  \mu_R
\hskip 0.2 cm ; \hskip 0.2 cm 
f(\mu)=f_L \hskip 0.1 cm {\rm for } \hskip 0.1 cm \mu_R< \mu < 1
\hskip 5 cm 
\eqno(9)
$$
By symmetry, $\mu _L=\mu _R$.  

In the region between the plates, there is a boundary in $\mu$ at $\mu_b(x)$
such that $\mu_b(-L)=-\mu_L$ and $\mu_b(L)=\mu_R$.
CR with $\mu>\mu _b$ are emitted by the left-hand plate, and CR with $\mu<\mu_b$ are emitted from the right-hand plate.
The function $\mu _b(x)$ is determined by whether or not CR are able to pass through the mirror  at $x=0$ where 
$B(0)=B_{\max}$. 
Hence $\mu_b(0)=0$, and conservation of magnetic moment determines $\mu_b$ at all other $x$ such that 
$$
\frac {1- \mu_b^2}{B(x)}=\frac {1}{B_{\max}}
\hskip 0.3 cm ; \hskip 0.3 cm
\mu _b(x)=\mp \left ( 1 - \frac {B(x)}{B_{\max}} \right )^{1/2}
\eqno(10)
$$
where $\mp \rightarrow -$ for $x<0$ and $\mp \rightarrow +$ for $x>0$.
The overall solution for the distribution function for all $\mu $ between the plates $-L<x<L$, 
 is
$$
f(\mu)=f_L \hskip 0.2 cm {\rm for } \hskip 0.2 cm x<0, \hskip 0.1 cm \mu > - \left (1 -\frac {B(x)}{B_{\max} } \right  )^{1/2}
$$
$$
f(\mu)=f_R \hskip 0.2 cm {\rm for } \hskip 0.2 cm x<0, \hskip 0.1 cm \mu < - \left (1 -\frac {B(x)}{B_{\max}  } \right )^{1/2}
$$
$$
f(\mu)=f_L \hskip 0.2 cm {\rm for } \hskip 0.2 cm x>0, \hskip 0.1 cm \mu > \left (1 -\frac {B(x)}{B_{\max} } \right )^{1/2}
$$
$$
f(\mu)=f_R \hskip 0.2 cm {\rm for } \hskip 0.2 cm x>0, \hskip 0.1 cm \mu <  \left (1 -\frac {B(x)}{B_{\max} } \right )^{1/2}
\eqno(11)
$$
From this it follows that
$$
\mu _L= \mu _R= \left (1 -\frac {B_{0}}{B_{\max} } \right )^{1/2}
\eqno(12)
$$
We choose a magnetic field with $x$-dependence
$B(x)=B_{0}+{ (B_{\max} -B_{0})}/{\cosh (x/h)}$ with
$B_{\max}=4B_{0}$
and  $h=L/4$ as plotted in figure 1.
This actually makes $B(L)=1.03B_0$, but this small departure from $B_0$ is inconsequential.

Equations (11) could be generalised to a CR distribution that is not monoenergetic by allowing $f_L$ and $f_R$ to
be functions of momentum $p$.

\begin{figure}
\includegraphics[angle=0,width=6cm]{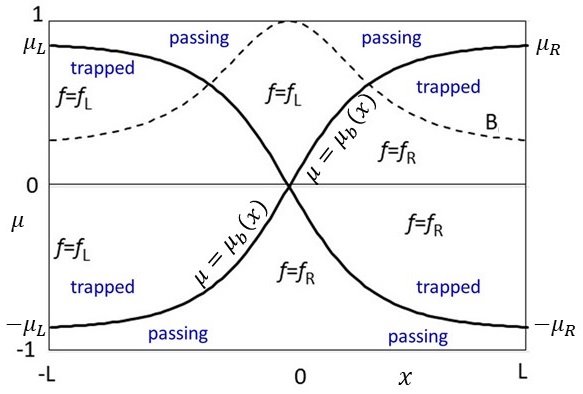}
\centering
\caption{
The  CR distribution function given by equations (11).
The full line is the boundary between populations of trapped and passing CR originating
at the left- and right-hand boundaries.
The magnetic field profile is given by the dotted line.
}
\label{fig:figure1}
\end{figure}

We take moments of the distribution function (equations (11)) to derive the density $n_L(x)$ and current density $ j_L(x)$ profiles.
The subscript $L$ (in $n_L$, $j_L$) denotes CR entering from the left. 
To avoid confusion on the meaning of $f$ for a monoenergetic distribution, we make $f$ a function of momentum $p$
but restrict the momentum range to $p_{\min} < p < p_{\max}$  where $\Delta p= p_{\max}-p_{\min}$, and $\Delta p$ is vanishingly small.
First we consider only CR originating at the left-plate by setting $f_R=0$.
The $x$-dependent number and current densities of the left-originating CR are
$$
n_L(x)=\int _{p_{\min}} ^{p_{\max}} \int _{\mu_b(x)} ^1  2 \pi f_L (p) p^2 dpd \mu
\hskip 8 cm
$$
$$
\hskip 1 cm 
= \frac {1}{2} 
\left [
1 
\pm\left (1 -\frac {B(x)}{B_{\max} } \right  )^{1/2}
\right ] 
\int _{p_{\min}} ^{p_{\max}}  4 \pi f_L (p) p^2 dp 
\hskip 6 cm
\eqno(13)
$$
where $\pm \rightarrow +$ for $x<0$, and $\pm \rightarrow -$ for $x> 0$, and 
$$
j_L(x)
= \frac {c}{4} \frac {B(x)}{B_{\max}}
\int _{p_{\min}} ^{p_{\max}}  4 \pi f_L (p) p^2 dp 
\hskip 6 cm
\eqno(14)
$$
where the $\pm$ sign disappears from the derivation of $j_L(x)$. The form of the expression for the current density
is the same for both $x<0$ and $x>0$.
As anticipated in section 2, $j_L(x)$ is proportional to $B(x)$.

Similar expressions can be derived for the density $n_R(x)$ and current density $j_R(x)$ of right-originating CR.
The total current density of CR entering and leaving the system at each of the end-plates, at $x=-L$ and  $x=L$, is the same:
$$
j_{\rm plate}=j(-L)=j(L)
= \frac {c}{4} \frac {B_{0}}{B_{\max}}
\int _{p_{\min}} ^{p_{\max}}  4 \pi \Big (f_L(p)-f_R(p) \Big ) p^2 dp 
\hskip 6 cm 
\eqno(15)
$$
The CR density at the plates is
$$
n_{\rm plate}=
\frac {1}{2}
\int _{p_{\min}} ^{p_{\max}}  4 \pi \Big ( f_L(p) +f_R(p) \Big ) p^2 dp 
\hskip 8 cm
$$
$$
\hskip 1 cm
\pm
\frac {1}{2} 
\left ( 1 - \frac {B_{0}}{B_{\max}} \right )^ {1/2}
\int _{p_{\min}} ^{p_{\max}}  4 \pi \Big ( f_L(p)- f_R(p) \Big ) p^2 dp 
\hskip 6 cm 
\eqno(16)
$$
where $\pm \rightarrow +$ at the left-hand plate and $\pm \rightarrow -$ at the right-hand plate.

The difference between the number densities at the left-hand and right-hand plates is
$$
\Delta n_{\rm plate}=
\left ( 1 - \frac {B_{0}}{B_{\max}} \right )^ {1/2}
\int _{p_{\min}} ^{p_{\max}}  4 \pi \Big ( f_L(p)  -f_R(p) \Big )p^2 dp 
\hskip 5 cm
\eqno(17)
$$
The relation between the current density and the difference between the plates in the number density is
$$
j_{\rm plate}= \frac {B_{0}}{B_{\max}}
 \left ( 1 - \frac {B_{0}}{B_{\max}} \right )^ {-1/2} 
\frac { c \Delta n_{\rm plate} }{4}
\hskip 8 cm
\eqno(18)
$$
This relationship can be made to look like a diffusion process
if we write, with gross approximation, that the density gradient is equal to the density difference
$\Delta n_{\rm plate}$ between the plates
divided by the distance $2L$ between the plates:
$$
\frac {\partial n}{\partial x} = -\frac {\Delta n_{\rm plate}}{2L}
\eqno(19)
$$
and define a mirror-induced diffusion coefficient $D_m$ as the constant in the diffusion equation
$$
j_{\rm plate}=-D_m \frac {\partial n}{\partial x} 
\eqno(20)
$$
With these definitions,
$$
D_m=
\frac {cL }{2}
\frac {B_{0}}{B_{\max}}
 \left ( 1 - \frac {B_{0}}{B_{\max}} \right )^ {-1/2} 
\eqno(21)
$$
Expressed in this way, $D_m \rightarrow \infty$ in a uniform magnetic field 
when propagation becomes ballistic, and the diffusion model breaks down.

$D_m$ can be compared to a representative Bohm diffusion coefficient
$D_B=cr_{g,\max}$
where $r_{g,\max}=p/eB_{\max}$ is the CR Larmor radius in the magnetic field at the centre of the mirror:
$$
\frac {D_m}{D_B} = \frac {1}{2}
\frac {B_{0}}{B_{\max}}
 \left ( 1 - \frac {B_{0}}{B_{\max}} \right )^ {-1/2} 
\frac {L}{ r_{g,\max}} 
\eqno(22)
$$
The mirror only acts as a mirror if $L$ is larger than the CR Larmor radius,
$L> r_{g,\max}$.  
Equation (22) suggests the possibility of sub-Bohm transport, $D_m<D_B$, if 
the magnetic field $B_0$ between mirrors (at the end-plates in our calculation) 
is much smaller than the magnetic field $B_{\max}$ in the mirror.
The ideal condition, $B_0\ll B_{\max}$, for strong mirror-inhibited transport
may be more easily satisfied by having small $B_0$ than by having large $B_{\max}$.

Another interesting feature of mirror-dominated transport is that 
the CR pressures are variable and anisotropic on the spatial scale of the mirrors.
This contrasts with diffusive transport where the CR pressure is smooth on the scale of diffusion length $D/u$.
This raises the possibility that mirror-induced CR pressure gradients may drive turbulence on the mirror scale which 
can be much smaller than the diffusion length $D/u$.
CR undergoing shock acceleration have large energy densities and pressures that can be 
as much as 10\% or
more of the hydrodynamic energy density $\rho u^2$.
Hence there can be large pressure differences across a mirror that might feed back onto the hydrodynamical structures of mirrors, possibly enhancing the mirrors and further inhibiting CR transport.
This can be seen in figure 2 which plots the spatial profiles of density, current density, and the parallel and perpendicular pressures
when CR originate only at the left-hand mirror ($f_R=0$).
$$
\hskip 1 cm
n(x)=
\left [
1 \pm\left (1 -\frac {B(x)}{B_{\max} } \right  )^{1/2}
\right ] 
 \frac {n_{\rm ref}}   {2} 
\hskip 8 cm$$
$$ \hskip 1 cm
j(x)=  \frac {B(x)}{B_{\max}} \frac {n_{\rm ref}c}{4}
\hskip 8 cm$$
$$\hskip 1 cm
P_\parallel (x)= 
\left [
1 \pm \left (1 -\frac {B(x)}{B_{\max} } \right  )^{3/2}
\right ] 
\frac {P_{\rm ref}}{2} 
\hskip 8 cm$$
$$\hskip 1 cm
P_\perp (x)=  
\left [
1 \pm  \left ( 1+ \frac {B(x)}{2B_{\rm max}}\right ) \left (1 -\frac {B(x)}{B_{\max} } \right  )^{1/2}
\right ]
\frac {P_{\rm ref}}{2} 
\hskip 8 cm
\eqno(23)
$$
where
$$
n_{\rm ref}=
\int _{p_{\min}} ^{p_{\max}}  4 \pi f_L (p) p^2 dp
\hskip .15 cm ; \hskip .15 cm
P_{\rm ref}=
\int _{p_{\min}} ^{p_{\max}} \frac {4 \pi}{3} f_L (p) cp^3 dp
\eqno(24)
$$

\begin{figure}
\includegraphics[angle=0,width=8cm]{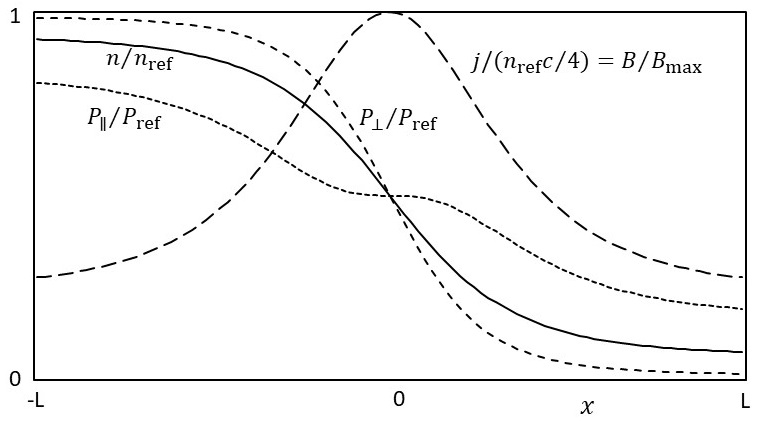}
\centering
\caption{
The density, current density, and perpendicular and parallel pressures corresponding to the
magnetic field in figure 1 when CR are injected only at the left-hand plate ($f_R=0$).
The reference quantities, $n_{\rm ref}$ and $P_{\rm ref}$, are given in equation (24).
}
\label{fig:figure2}
\end{figure}

\section{An illustrative model for CR shock acceleration between converging mirrors in the absence of small-angle scattering}

Consider a configuration in which two ideal mirrors upstream and downstream of a parallel shock 
(magnetic field aligned with the shock normal) are represented by 
perfectly reflecting plates.
In this model, the magnetic field is uniform between the reflecting plates acting as mirrors.
This model is far from any realistic astrophysical situation,
but it clarifies salient aspects of shock acceleration when CR are reflected between the upstream
and downstream plasmas by mirrors each side of the shock.
In particular it shows that significant acceleration is only possible if
the upstream plasma contains multiple semipermeable mirrors.

Initially at time $t=0$ the upstream plate is at a distance $L_0$ from the shock
 and moves towards the shock at velocity $u$ in the shock rest frame.
The downstream plate is initially coincident with the shock and moves away from the shock at a velocity $u/4$.
After time $t$ the distance between the two mirrors is $L=L_0-3ut/4$.

A CR has momentum ${\bf p}$ with a component $p_\parallel$ parallel to the shock normal and consequently also parallel to the magnetic field.
$p_\parallel$ increases by 
$
\Delta p_\parallel =  (3 u/2c)p
$
on each cycle between upstream and downstream taking a time 
$
\Delta t = (2L/c) (p/p_\parallel)
$
for each cycle.
In the limit of $u\ll c$ and small $\Delta p_\parallel$ and $\Delta t$,
$$
\frac {dp_\parallel}{dt}=\frac {3}{4} \ \frac {u p_\parallel }{L_0-3ut/4}
\eqno(25)
$$
$p^2=p_\perp^2+p_\parallel^2$.  
The magnitude of the perpendicular part of the momentum, $p_\perp =p_0 \sqrt {1-\mu_0^2}$, is constant for perfect mirrors and no small-angle scattering.
$p_0$ and $\mu _0$ are the initial values of $p$ and $\mu$.
Accordingly, $\mu $ and $p$  after time $t$  are
$$
\mu = \mu _0 
\left [
 \mu _0^2  +
  (1- \mu _0^2)\  \left (1-\frac {3ut}{4L_0 }   \right)^2  
\right ]^{-1/2}
{\rm and} \hskip .2 cm
\frac {p}{p_0}= 
\left [ \frac {1-\mu_0^2}{1- \mu ^2} \right ]^{1/2}
\eqno(26)
$$
giving
$$
\frac {p}{p_0}= 
\left [
1-\mu _0^2+\mu _0^2   \left (1-\frac {3ut}{4L_0 }   \right)^{-2}   
\right ] ^{1/2}
\eqno(27)
$$
Ignoring all other effects, acceleration terminates when  the upstream  mirror passes through the shock at time $t=L_0/u$.
On termination,
$$
\mu = \frac {4 \mu_0}{(1+15 \mu_0^2 )^{1/2}}
\hskip .2 cm {\rm and} \hskip .2 cm
\frac {p}{p_0} = 
(1+15 \mu_0^2)^{1/2}
\eqno(28)
$$
However, the mirrors are transparent for CR with $\mu > (1-B_{\rm trap}/B_{\rm mirror})^{1/2}$
in which event the CR escapes the mirror when its momentum reaches
$$
p_{\rm escape}=p_0 \left [ \frac {B_{\rm mirror}}{B_{\rm trap}}\ (1-\mu _0^2) \right ]^{1/2}
\eqno(29)
$$
where  $B_{\rm mirror}$ and $B_{\rm trap}$ are the magnetic fields at and between the mirrors respectively.
The momentum gain from the converging mirrors is the lower value of that imposed by (i) the upstream mirror passing through the shock,
and (ii) the CR escaping through the mirror:
$$
\frac {p}{p_0} = \min
\left \{
(1+15 \mu_0^2 )^{1/2}
\ ,\ 
(1-\mu _0^2) ^{1/2}
\left ( \frac {B_{\rm mirror}}{B_{\rm trap}} \right )^{1/2} 
\right \}
\eqno(30)
$$
The energy gain is plotted in figure 3 which shows that an energy increase by a typical factor of 2 is reasonable. 
Multiple trapping episodes are needed if the CR energy is to increase by an order of magnitude or more.

A more realistic calculation will be needed to allow for changes in the magnetic moment due to
small-angle scattering by turbulence between the mirrors and at shocks which are oblique rather than parallel.
Scattering in angle would diffuse the CR in $\mu$ and counteract the otherwise monotonic increase in $\mu $ during acceleration.
This would allow some CR to remain in
the accelerating region for a longer period of time without escaping through the mirror.

It should be noted that the shock compresses the component of the magnetic field that is perpendicular
to the shock normal.
This will increase the magnetic field at downstream mirrors 
with the consequences that the downstream mirrors are stronger than the upstream mirrors and that 
CR are more likely to escape upstream than downstream.

\begin{figure}
\includegraphics[angle=0,width=7cm]{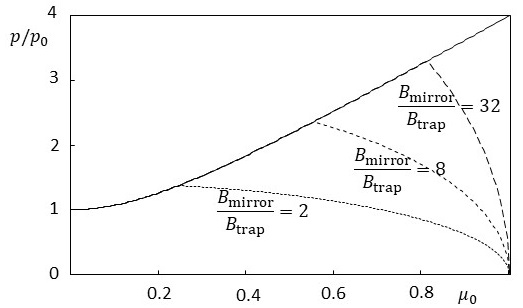}
\centering
\caption{
Energy (momentum) increase as a function of the initial $ \mu _0$
for different mirror strengths ($B_{\rm mirror}/B_{\rm trap}$).
No small-small angle scattering.
}
\label{fig:figure1}
\end{figure}
\section{A Vlasov-Fokker-Planck (VFP) model for transport in static mirror fields with small-angle scattering}

An ideal next step would be to solve the full momentum-dependent transport equation for CR
as they cross and recross a shock embedded in an evolving turbulent magnetic field which is structured both on a
sub-Larmor scale causing small-angle scattering and on a super-Larmor scale causing mirroring.
Such a calculation is beyond the scope of the present paper.
Here we limit ourselves to a calculation of CR transport in a two-dimensional ($x,y$) stationary plasma
in which the magnetic field is unchanging.
The magnetic field
consists of a uniform field $B_0$ in the $x$ direction to which components $B_x(x,y)$ and $B_y(x,y)$
are added to locally vary the direction and magnitude of the field lines.
(Note: the $x$ direction is fixed in this section, unlike in section 2 in which $x$ followed the field lines).
The aim of the calculation is to assess whether mirroring and trapping can inhibit CR transport
to levels comparable with small-angle Bohm diffusion and thereby provide a route for CR acceleration to,
and conceivably beyond, the Hillas limit.

The crucial point of comparison is that the DSA acceleration timescale is $D/u^2$ which is the time taken 
in diffusion theory for the shock
to overtake a CR precursor of height $D/u$ (Lagage \& Cesarsky 1983a,b).
We test whether mirroring can, by inhibiting CR transport, reduce the precursor scaleheight
and thereby increase the acceleration rate and the energy to which CR can be accelerated.

In this and subsequent sections we present solutions to the VFP equation for 
the transport of monoenergetic CR in a pre-determined static magnetic field.
We include small-angle scattering as an additional term in the VFP equation so that we can compare confinement by mirrors  with confinement by mixtures of scattering and mirroring.

An inherent assumption in our model is that the magnetic field can be separated into super-Larmor structures modelled on a spatial grid
and sub-Larmor structures modelled as sub-grid damping of CR anisotropy.
Maybe the most interesting case would be the intermediate case in which structures on the Larmor scale
are included, but that would require a more complicated calculation, and we find that much can be learned from the present model.

The non-zero components of the magnetic field, $B_x$ and $B_y$, 
can be represented as the curl of a vector potential $A_z$ in the $z$ direction.
We impose $B_z=0$, and $A_x=A_y=0$.
$B_y$ is set to zero at each boundary in $x$. 

CR are initially placed in a small region close to the left-hand reflective boundary at $x=0$.
No further CR are injected during the calculation.
CR reaching the right-hand boundary in $x$ are allowed to escape freely without reflection.

Because the magnetic field is constant in time there is no electric field and the energy of each CR is constant.
The imposed small-angle scattering is constant in space and time.
We perform calculations with a range of different scattering rates.

In our approximation that the magnetic field varies on scales much larger than the Larmor radius, CR trajectories are confined to field lines.
Trajectories are the same for CR of any energy.
Gyration about field lines removes anisotropies across the magnetic field.
In contrast to cross-field transport, which is negligible and neglected, CR are free to move
along field lines with a degree of anisotropy that can reach beam-like levels.


We adopt a formalism, often used in solution of the VFP equation,  in which the CR distribution function is expanded in spherical harmonics
$$
f=\sum_{n=-\infty}^{\infty}\sum _{m=0}^{| n |} f_n^m ({\bf r}) P_n^m (\mu  ) e^{im\phi}
\hskip .2 cm {\rm where} \hskip .2 cm
f_{-n}^m ({\bf r})= \left ( f_n^{m} ({\bf r}) \right )^*
\eqno(31)
$$
The equations for the evolution of the coefficients $f_n^m$ in 2D $x,y$ space are given in Bell et al (2006); see also Reville \& Bell (2013).
Our case is simpler than the general case considered by Bell et al (2006) because the electric field is zero, the distribution is monoenergetic,
 and the only collision term we consider is small-angle scattering.
However, we complicate the evolution equations by aligning the $\theta =0, \ \mu =1$ axis with the local magnetic field.
The benefit of this move is that all $m>0$ terms can be ignored in the limit in which
CR follow the field lines.
The $m=1$ terms describe drifts across the magnetic field, but these are small when the Larmor radius is small compared with other 
scalelengths.
We keep only the $m=0$ terms, in which case $f(x,y,\mu)$ is an expansion in Legendre polynomials alone
with 
 $\mu={\bf b}.{\bf v}/|{\bf v}|$ where 
${\bf b}={\bf B}/|{\bf B}|$.
$$
f({\bf r},\mu)=
\sum _n f_n({\bf r}) P_n (\mu)
\eqno(32)
$$ 
In these co-ordinates, and neglecting cross-field drifts and higher order cross-field anisotropies,  the VFP equation takes the form
$$
\sum_n \frac {\partial f_n}{\partial t} P_n(\mu)
+
\sum_n 
{\rm v} \left ( \nabla _\parallel  f_n \right )  \mu P_n(\mu)
+
\sum_n 
\left ( {\bf v}.\nabla \mu \right )  f_n  \frac {\partial P_n(\mu)}{\partial \mu}
\hskip 10 cm
$$
$$
\hskip 3 cm 
=
-\nu \sum _n \frac {n(n+1)}{2} f_n P_n (\mu)
\eqno(33)
$$
This equation simplifies when gyro-averaged
since any component of ${\bf v}$ perpendicular to $\bf B$ averages to zero, and its square 
averages to 
 ${\rm v}_\perp ^2/2$.
Using 
${\bf b}.[ ( {\bf b}.\nabla ) {\bf b }]=0$, 
gyro-averaging yields
$$
{\bf v}.\nabla \mu 
\rightarrow
 \frac {{\rm v}_\perp ^2}{2{\rm v}B} \nabla _\perp . {\bf B}_\perp 
\eqno{(34)}
$$
From $\nabla. {\bf B}=0$, 
$\ \nabla _\perp . {\bf B}_\perp= - \nabla _\parallel B_\parallel $
where the subscripts  $_ \perp $ and  $_ \parallel $ 
refer to components perpendicular to and parallel to the magnetic field at the point of interest.
Consequently, the VFP equation takes the form
$$
\sum _n \bigg \{
\frac {\partial f_n}{\partial t} P_n(\mu)
+     \mu P_n(\mu) {\rm v} \nabla _\parallel  f_n
- \frac {\nabla _\parallel B}{2B}  (1-\mu ^2) \frac {\partial P_n(\mu)}{\partial \mu} {\rm v}  f_n
\hskip 5 cm $$
$$
\hskip 4 cm
+ \nu \frac {n(n+1)}{2} f_n P_n (\mu)
\bigg \}
=0
\hskip 12 cm
\eqno(35)
$$
This is the equivalent of equation (6) in section 2.
The corresponding equations for the evolution of the coefficients $f_n$ are
$$
\frac {\partial f_n}{\partial t}=
- \frac {n}{2n-1} {\rm v} 
{\bf b}.\nabla f_{n-1}
- \frac {n+1}{2n+3}{\rm v} 
{\bf b}.\nabla f_{n+1}
- \frac {n(n+1)}{2} \nu f_n
\hskip 6 cm
$$
$$
\hskip 2 cm 
+\frac {{\rm v}}{2|{\bf B}|} {\bf b}.\nabla |{\bf B}|
\left (
\frac {(n+1)(n+2)}{2n+3} f_{n+1}
- \frac {n(n-1)}{2n-1} f_{n-1}
 \right )
\hskip 6 cm
\eqno(36)
$$

\section{A candidate site for mirror-dominated transport: Cassiopeia A }

The above VFP equation might be solved for many different problems in and beyond astrophysics.
With appropriate scaling, solution of the same VFP equation might be relevant to UHECR transport in the intergalactic medium or CR transport in the interstellar medium, or to energetic particles in the solar system.
We could solve a representative problem in dimensionless units, but we prefer to adopt a particular application and solve
the VFP equation in dimensional units.
The particular application we choose is the transport of PeV protons,
which may or may not exist, ahead of the outer shock of the iconic supernova remnant Cassiopeia A (Cas A).

The approximately 340-year-old SNR, Cas A,
is the brightest extra-solar radio source in the sky, comparable with Cygnus A.
The radio emission is generated by synchrotron-emitting energetic electrons,
 so there is good prima facie reason to suppose that Cas A might be a good accelerator of high energy CR.
However, it has been found that its gamma-ray spectrum steepens at around 10TeV (Ahnen et al 2017, Abeysekara 2020), indicating a turnover in the CR spectrum at around 100TeV.
Consistent with results from MAGIC and VERITAS, recent results from LHAASO impose upper limits on CR densities in Cas A at energies up to 1 PeV (Cao et al 2024).
Gamma-ray emission depends on the presence of a dense target for the CR-background interaction.
As Cao et al point out, any substantial population of PeV CR produced now or earlier in the evolution of Cas A would need to be outside the dense shell at present.

A turnover in the CR spectrum at 100TeV accords well with theories of magnetic field amplification and DSA.
 Zirakashvilii \& Ptuskin (2008) and Bell et al (2013) showed that 
 CR above 100-200TeV are unable to amplify the Larmor-scale
magnetic field needed for acceleration beyond this energy.
Our aim here is to explore MDSA as a process by which CR might be accelerated to, or possibly beyond, 1PeV.

In MDSA, a combination of mirroring and angular scattering may confine CR relatively close
to the shock and allow acceleration to continue beyond 100TeV.
Because MDSA depends on the presence of mirror-inducing magnetic field structures, it can be expected to be spasmodic and less effective than standard DSA, leading
to a steepening of the gamma-ray spectrum which would be in accordance with observations.
As discussed above, magnetic mirrors are strengthened as they pass through the shock and CR preferentially escape upstream
instead of being carried downstream into the heart of the SNR.
Gamma-ray emission by PeV CR will therefore be weakened by the lack of target material with which to interact.  
This would further contribute to a reduction in gamma-ray emission that would otherwise indicate their presence.

Cas A is a good candidate for MDSA since the plasma upstream of the shock is strongly disturbed by high velocity, high density, knots penetrating beyond
the shock and, maybe more importantly, by the presence of dense disordered circumstellar material thrown off by the pre-supernova
(Arias et al 2018, Vink et al 2022, Milisavljevic et al 2024).
The strong activity seen in Cas A is probably caused by the shock passing through a dense shell of
circumstellar material.
Consequently there is good reason to suppose that the magnetic field ahead of the shock is far from uniform
with the likely result that mirroring may restrict CR transport during acceleration by the outer shock.
The question is whether mirroring can confine CR close to the shock,
but not confine them so well that CR get lodged in magnetic traps which are overtaken by the shock.
Or equivalently, viewed in the shock rest frame, CR acceleration would be terminated at low CR energy if strong mirroring
prematurely 
carries CR  through the shock and away downstream.

\section{Solution of the VFP equation with parameters based on Cas A }

We now solve the VFP equation for CR transport in a 2D stationary plasma with no hydrodynamic motion
and a non-evolving magnetic field as described in section 5.
The calculation can be thought of as loosely applying to a region immediately ahead of the outer shock in Cas A.
The left-hand  boundary in our figures at $x=0$ is reflective to represent strong mirroring at or immediately behind a shock.
The shock velocity in Cas A is of the order of 60 times smaller than the velocity of the CR,
so it does not concern us unduly that the plasma is stationary with respect to the boundary.
An improved model, work for the future, would have the CR and the  upstream magnetic field flowing into and passing through a shock.
The improved model would include CR angular scattering at the shock and in the immediately downstream plasma
where the magnetic field is compressed.

In our calculation, the right-hand boundary in $x$ at a distance of $L_x=5\times 10^{16} {\rm m}$ (=1.7parsec, 
approximately the radius of the shell of Cas A) 
is a free escape boundary to represent 
CR being lost to the surrounding interstellar medium (ISM).
The boundaries in $y$ are periodic and similarly separated by $L_y=5\times 10^{16} {\rm m}$
for calculations with a non-uniform field.
For calculations with a uniform magnetic field there is no dependency on $y$ and we set $L_y=0.05\times 10^{16} {\rm m}$
for more rapid computation.

The magnetic field is ${\bf B}= \nabla \times \left ( A_z \hat {\bf z} \right )$.
The magnetic field consists of a uniform component ($B_0$) in the $x$-direction 
and additional non-uniform components in the $x$ and $y$ directions. 
We choose $B_0=50 \mu {\rm G}$ which is intermediate between a field of $3\ {\rm to} \ 30 \mu {\rm G}$ 
expected in an undisturbed ISM (Beck 2015) and the field of  $100$ to $500 \mu {\rm G}$ 
observed at the shock (Vink \& Laming 2003, V\"{o}lk et al 2005).

For the benefit of easy comparison we use the same structure for the non-uniform part of the magnetic field in every calculation.
$$
A_z=
B_0 y+
\alpha _B  B_0 L_y
\sum _{j,k}
 \cos \left ( \frac {j \pi x}{L_x} \right ) 
\left ( A^c_{jk}\cos \left (\frac {k \pi y}{L_y} \right )+A^s_{jk}\sin \left (\frac {k\pi y}{L_y} \right ) \right )
\eqno(37)
$$
where $k$ is even but $j$ takes both odd and even values.  
$A^c_{jk}$ and  $A^s_{jk}$ are chosen by a random number generator and scaled by $(j^2+k^2)^{-1/2}$.
Additionally, the non-uniform part of $A_z$ is artificially reduced by an ad hoc multiplier close to the boundaries in $x$.
Without this, the amplitude of the  turbulent field is greater at the $x$ boundaries due to the in-phase (or anti-phase) addition of 
all its $\cos(j\pi x/L_x)$ harmonic components;
components of the form  $\sin(j\pi x/L_x)$   are disallowed by the reflective boundary conditions.
The field-lines and the magnitude of the magnetic field in the reference calculation
($\alpha _B=0.1$ in equation (37)) are plotted in figure  4.

CR are initialised as a monoenergetic isotropic population within 0.05parsec of the left-hand $x$ boundary.
No further CR are injected during the course of the calculation.
The calculation is run for 50 years, during which time a shock with velocity $5000{\rm km \ s}^{-1}$ would advance
0.25parsec.
In the reference calculation, the angular scattering rate is set to $0.01 \nu_{\rm B}$
where we define $\nu_{\rm B} = \omega _g/2\pi$ and
$\omega _g$ is the Larmor frequency of a PeV proton in the magnetic field $B_0$.
Defined in this way, 
$\nu_{\rm B}$ represents Bohm-like exponential decay of the $f_1$ current anisotropy in one Larmor gyration.
The expansion in  Legendre polynomials is extended to 40th order for calculations with the lowest
small-angle scattering rate ($\nu=0.001 \nu _{\rm B}$).
For calculations with larger scattering rates an expansion to 20th Legendre polynomial is sufficient.
The number of spatial gridpoints is 500 in each of the $x$ and $y$ dimensions, corresponding to a cell size of $10^{14}{\rm m}$.

Figure 5 plots the first three coefficients, $f_0,\ f_1,\ -f_2$,  of the Legendre polynomial expansion after 50 years.  
The normalisation of each coefficient is the same in each case.  
Noticeably, $|f_2|$ can be larger than $f_0$, although not reaching the maximum allowed value of $5f_0$
for a perfectly beamed  distribution.
\begin{figure}
\includegraphics[angle=0,width=7cm]{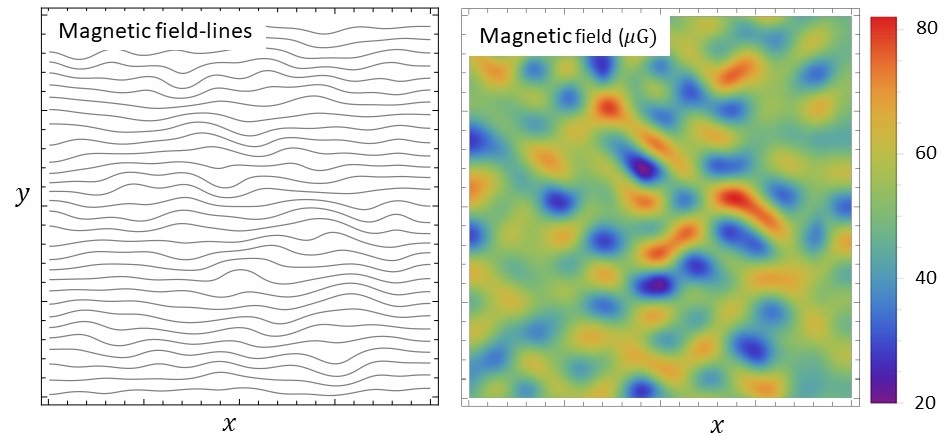}
\centering
\caption{
Magnetic field lines (left) and the magnitude (right) of the magnetic field.
As in all our spatial plots, the spatial box is $L_x$ (horizontal) by $L_y$ (vertical).
}
\label{fig:figure4}
\end{figure}
We perform calculations with the same magnetic field but with four different small-angle scattering rates, 
$\nu=\nu _{\rm B}$, $\nu=0.1 \nu _{\rm B}$, $\nu=0.01 \nu _{\rm B}$ (the reference calculation), and $\nu=0.001 \nu _{\rm B}$.
Figure 6 plots $f_0$ for each of these cases.
The numerical normalisation of $f$ is the same throughout in figures 5 to 7.
The $\nu = 0.01 \nu _B$ data is the same in both figures 5 and 6.
\begin{figure*}
\includegraphics[angle=0,width=11.5cm]{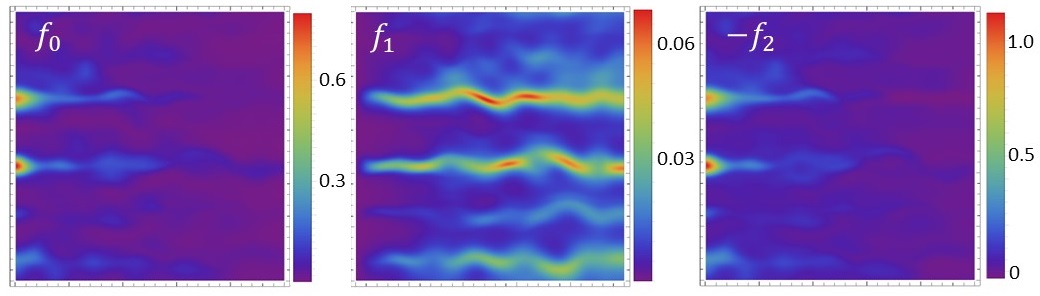}
\centering
\caption{
The CR distribution after 50 years in the reference calculation ($\nu =0.01 \nu_B$). 
Plots, from left to right, of the isotropic part $f_0$, the current component $f_1$,
and $-1$ times the 2nd-order tensor component $f_2$.
The CR were initially placed close to the left-hand reflective boundary, and freely escape at the right-hand boundary.
The boundaries at the top and bottom are periodic.
}
\label{fig:figure5}
\end{figure*}
\begin{figure*}
\includegraphics[angle=0,width=15.25cm]{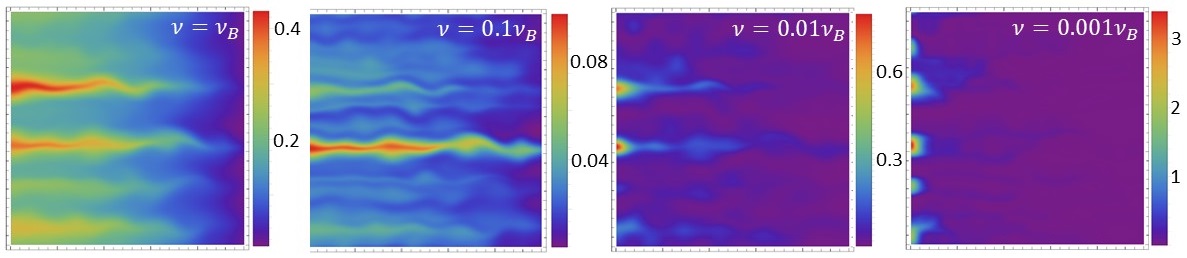}
\centering
\caption{
The isotropic $f_0$ components of the CR distribution with a non-uniform magnetic field after 50 years for different
levels of small-angle scattering
$\nu = \nu _{B}$, $\nu =0.1 \nu _{B}$, $\nu = 0.01 \nu _{B}$, $\nu = 0.001 \nu _{B}$.
The plot for $\nu = 0.01 \nu _{B}$ is the same as the plot of $f_0$ in figure 5.
}
\label{fig:figure6}
\end{figure*}
\begin{figure*}
\includegraphics[angle=0,width=15.5cm]{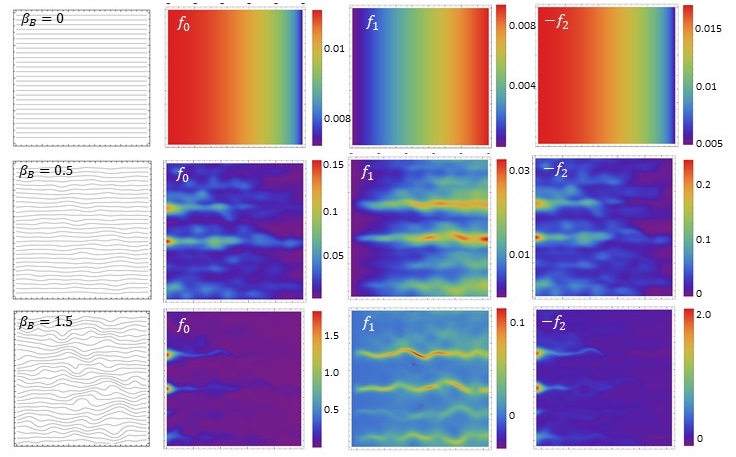}
\centering
\caption{
The $f_0$, $f_1$ and $-f_2$ components of the CR distribution after 50 years with varying degrees of magnetic field non-uniformity
corresponding to $\beta _B=0$ (uniform field, no non-uniformity), $\beta_B=0.5 $ (non-uniform component reduced by 50\%),
and $\beta _B=1.5$ (non-uniform component increased by 50\%).
The reference case of $\beta _B=1$ can be found in figure 5.  In each plot, $\nu =0.01 \nu _B$ (the reference value for 
small-angle scattering).
}
\label{fig:figure6}
\end{figure*}
We also repeat calculations with our reference scattering rate, $\nu=0.01 \nu _B$, 
but with different magnetic field non-uniformities, including the case of a uniform magnetic field.
These are plotted in figure 7 in which the magnetic fields are
set by multiplying the non-uniform part of the magnetic field ($\alpha _B=0.1$) by a further factor $\beta _B$ 
where $\beta _ B =0,\ 0.5, \ {\rm and}\  1.5$.
In the case of a uniform magnetic field  ($\beta  _B=0$) the problem is one dimensional and the
CR profiles are smooth.
Even a relatively small magnetic   non-uniformity  ($\beta _B=0.5$) produces a significant
non-uniformity in the CR number density which translates into non-uniform CR pressure profiles.
As noted above in section 3, since CR may be accelerated with efficiencies of 10\% or more,
the consequently large CR pressure gradients
may be sufficient to drive or amplify turbulence in the plasma upstream of a SNR outer shock.
This may suggest a fruitful future inquiry into the evolution of the turbulence.

\section{Mirror-dominated transport inhibition and shock acceleration by SNR}

The scaleheight of the CR precursor ahead of a SNR shock increases with CR energy because the CR Larmor radius increases with energy.
CR at low energy are accelerated rapidly because their precursor scaleheight
is much smaller than the SNR shock radius.
At higher energies, acceleration is slower because the scaleheight is larger.
Acceleration terminates when
 the scaleleheight approaches the radius of the SNR.

Effective acceleration by a steadily propagating shock depends on the comparison of two lengths: 
(i) the scaleheight $h$ of the CR precursor 
(ii) the distance $L_{\rm shock}=ut$ propagated by the shock.
Acceleration is effective if $h<L_{\rm shock}$.

Our calculation does not include a shock,
and the left-hand boundary does not advance into the plasma, so we cannot directly assess the competition between
the advance of a shock and the advance of the CR precursor scaleheight.
However, we can compare the advance of the CR precursor with the distance a shock would advance if it were included in the model.

In a diffusive system $h\propto t^{1/2}$, while a shock would advance linearly with time, $L_{\rm shock }\propto t$.
Initially $h$ increases more rapidly than $L_{\rm shock}$, but a time $t_{\rm overtake}$ is reached at which $L_{\rm shock}$
overtakes $h$.
When applied to SNR, CR are accelerated if $t_{\rm overtake}$ is less than the age of the SNR.

In a mirror-dominated system, the time dependence of $h$ does not have a simple analytic expression,
but we can still compare the distance travelled by CR with the distance a shock would travel.
Our procedure is to calculate the fraction of the initially injected CR that
remain within a distance $L_{\rm shock}$ ($\propto t$) of their point of origin close to $x=0$.

We define the function $F_{\rm shock}(t)$ (as plotted in figure 8) to represent the number of CR 
within the distance $L_{\rm shock}=ut$ that a shock would propagate:
$$
F_{\rm shock}(t)
=\int _0^{L_y} \int _0^{L'_{\rm shock}} f_0 dx dy
\hskip .2 cm {\rm where} \hskip .2 cm
L'_{\rm shock}=\max(h_0,ut)
\eqno(38)
$$
$h_0$ is set to the distance (0.05parsec) in  which CR are contained at $t=0$.
The ratio $F_{\rm shock}(t)/F_{\rm shock}(0)$ is the fraction of CR  remaining at time $t$ within 
the shock-propagation distance from the left-hand boundary.
While acknowledging that the real situation is much more complicated, we use this 
to compare CR confinement in different combinations of mirroring and small-angle scattering.

The plots of  $F_{\rm shock}(t)/F_{\rm shock}(0)$, in figure 8, with and without magnetic non-uniformity, show the effect of
mirroring for different rates of small-angle scattering.
The curves in figure 8 have a characteristic structure of an initial decrease in $F_{\rm shock}(t)/F_{\rm shock}(0)$
while the CR precursor is advancing more quickly than the shock.
In diffusion theory this is the equivalent of $h$ increasing more rapidly than $L_{\rm shock}$.
At later times, $F_{\rm shock}(t)/F_{\rm shock}(0)$ increases, or decreases less rapidly, when the opposite condition pertains
as the shock overtakes the CR precursor.

In three of the plots of $F_{\rm shock}(t)/F_{\rm shock}(0)$ in figure 8 ($ \nu =1,0.1,0.001 \nu_B$) we compare the standard magnetic field ($\beta _B=1$)
with a uniform magnetic field ($\beta _B =0$).
For the standard reference scattering rate  ($\nu = 0.01 \nu_B$) we also plot results for $\beta _B=0.5$ and $\beta _B=1.5$.

Non-uniformities in the magnetic field make little difference to  $F_{\rm shock}(t)/F_{\rm shock}(0)$ when $\nu=\nu_{\rm B}$
since small-angle diffusion dominates trapping by mirrors.
Bohm scattering rapidly damps any anisotropy, and the pitch angle is randomised in the time taken to travel
the scalelength of magnetic field variation.

The magnetic uniformities have greatest impact when scattering is small ($\nu = 0.001 \nu _{\rm B}$)
and mirroring is allowed to dominate.
When  $\nu = 0.001 \nu _{\rm B}$, CR have a mean free path $80 \times $ larger than the size of the computational box.  
As expected for a uniform magnetic field and nearly unscattered propagation of an initially isotropic distribution,
the CR density evolves to become spatially uniform with a density that is proportional to $1/t$.
The shock propagation distance $L_{\rm shock}$ increases in proportion to $t$.
This accounts for the time-independent flat horizontal line for $F_{\rm shock}(t)/F_{\rm shock}(0)$
in figure 8 for 
 $\nu = 0.001 \nu _{\rm B}$ and $\beta _B=0$.

Trapping appears particularly effective when  $\nu = 0.001 \nu _{\rm B}$ and $\beta _B=1$.
However, the spatial plot for these parameters in figure 6
shows that the trapped CR stay close to the left-hand boundary where they would rapidly be overtaken by a shock
and be carried away downstream with limited acceleration as considered in section 4.

According to figure 8, increasing the scattering from $\nu =0.001 \nu_{\rm B }$ to $\nu =0.01 \nu_{\rm B }$ reduces the confinement,
but it may be more conducive to acceleration because CR are not as strongly trapped
and are less likely to be carried away into the downstream plasma.
A small amount of scattering allows CR to pass through mirrors.
If the escape rate is matched to the advance of the shock, this may allow some CR to stay with the shock for longer
and gain more energy.

Increasing the scattering rate to $\nu =0.1 \nu_{\rm B}$ further reduces the trapping, thus reducing the confinement,
but also allowing the CR to travel further towards the right-hand boundary.
A more sophisticated calculation, including a travelling shock, is needed
to discover the optimal level of scattering for CR acceleration.
 
Overall, our calculations suggest that a suitable combination of magnetic non-uniformity and small-angle
scattering can produce Bohm-level confinement without the need for Larmor-scale structure in the
magnetic field.
Of all the curves plotted in figure 8, the case of $\nu =0.01 \nu_{\rm B }, \  \beta_B=1.5$ 
in the top-right plot in figure 8 appears most suitable for CR acceleration.
The form of the curve resembles that in the top-left plot for Bohm diffusion ($\nu=\nu_B$) 
in a uniform field   ($\beta _B=0$).
$F_{\rm shock}(t)/F_{\rm shock}(0)$ is smaller in other plots in figure 8, but not always by a large margin.

We have chosen configurations in which the magnetic field varies on scales characteristically  
$10\times$ larger than the CR Larmor radius ($6.7 \times 10^{14}{\rm m}$).
This relatively large non-uniformity scalelength, compared with the Larmor radius, has been needed to justify the
assumption that CR are tied to magnetic field lines.
A reduction in the magnetic scalelength would be expected to increase confinement,
further inhibit transport, and accelerate CR more rapidly and to higher energies.
The important regime of magnetic non-uniformity
on a scale slightly greater than the Larmor radius will need an improved computational model that includes
cross-field drifts and $m>0$ anisotropies.
Lemoine (2022, 2023) has previously demonstrated the importance of this theoretically and computationally challenging regime for
CR  transport and second order Fermi acceleration.

\begin{figure}
\includegraphics[angle=0,width=8.55cm]{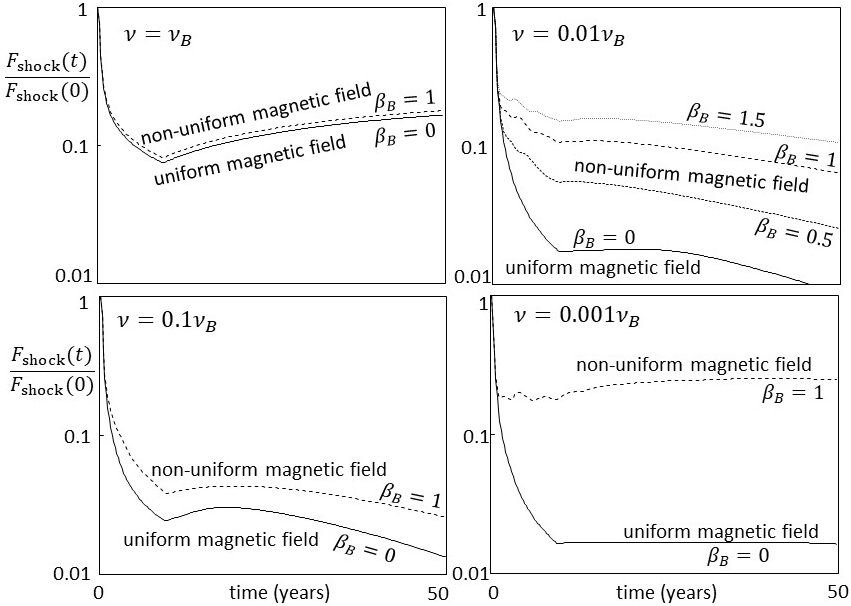}
\centering
\caption{
$F_{\rm shock}(t)/ F_{\rm shock}(0)$  with uniform ($\beta _B=0$) and non-uniform magnetic field ($\beta _B=1$), for different
small-angle scattering rates, $\nu = \nu _{B}$, $\nu =0.1 \nu _{B}$, $\nu = 0.01 \nu _{B}$, $\nu = 0.001 \nu _{B}$.
Additionally for the reference scattering rate ($\nu = 0.01 \nu _{B}$), results are included for 
the non-uniformity in the magnetic field increased by 50\% ($\beta _B=1.5$) and reduced by 50\% ($\beta _B=0.5$).
}
\label{fig:figure5}
\end{figure}

\section{A generalised Hillas limiting energy}

The Hillas energy, $E=uBR$, for CR acceleration by DSA can be derived by assuming that 
the minimum CR diffusion coefficient is $D_{\rm Bohm} =r_gc$ and that the acceleration time is $h/u$
where the CR precursor scaleheight is $h=D_{\rm Bohm}/u$.
A different derivation of the maximum CR energy is needed 
if mirroring is important.

Suppose that CR are confined in traps of characteristic length $L_{\rm trap}$ which is the distance between
the mirrors that control entry and exit from a trap.
Suppose that a CR passes from one trap to an adjacent trap with a characteristic escape probability $q_{\rm trap}$.
Omitting numerical factors of order 1, the CR performs a random walk of steplength  $L_{\rm trap}$
and timestep  $L_{\rm trap}/c q_{\rm  trap}$.
The resulting order-of-magnitude diffusion coefficient 
($ {\rm steplength}^2/{\rm timestep}$)
is 
$$
D_{\rm trap}=cL_{\rm trap}q _{\rm  trap}
\eqno{(39)}
$$
The corresponding precursor scaleheight ahead of the shock is
$h_{\rm trap}= (c/u)L q_{\rm trap}$,
and the CR acceleration time is $\tau_{\rm trap}=q_{\rm  trap} cL_{\rm trap}/u^2$.
If $R$ is the spatial extent of the system, the Hillas condition is determined by $\tau_{\rm trap}< R/u$.
This generalised Hillas condition can be written in the form
$$
E ({\rm in \ eV}) < \frac {uBR}{q_{\rm  trap}} \left (\frac {r_g}{L_{\rm trap}} \right ) 
\eqno{(40)}
$$
where
$B$ is a characteristic field in mirrors and $r_g$ is the Larmor radius in the same field.
Note that $B$ cancels from equation (40) since $r_g$ is inversely proportional to $B$.
Comparison of equation (39) with equation (21) in section 3, suggests that a
characteristic escape probability from the trap is
$$
q_{\rm trap}=
\frac {B_0}{2B_{\max}} \left ( 1 - \frac {B_0}{B_{\max}} \right )^{-1/2}
\eqno{(41)}
$$
Figure 9 is a plot of the ratio of the maximum CR energy to the Hillas energy $E_{\rm Hillas}=uBR$ 
as defined by combining equations (40) and (41) when
$L_{\rm trap}/r_g=10$.
For the parameters adopted for the VFP calculation in section 7
($u=5000 {\rm kms}^{-1}$, $B=50 \mu{\rm G}$, $R=1.7{\rm parsec}$), the Hillas energy 
$E_{\rm Hillas}$ is 1.3PeV.

The analysis in section 3, and hence this estimate of $q_{\rm trap}$, starts from the assumption that CR 
emerging from the left-hand and right-hand plates are isotropic.
The left-hand and right-hand plates in section 3 are the equivalent of the mid-points between mirrors. 
This assumption of isotropy between mirrors may be incorrect since CR with large $| \mu |$ at the midpoints
escape through the mirrors,  while CR with $ \mu$ close to zero are confined in the trap.
The transfer of CR between trapped and untrapped trajectories requires a more sophisticated calculation.

Taken at face value, equation (40) suggests that the highest energy is reached if $q_{\rm trap}$
is indefinitely small.
However, $q_{\rm trap}$ must not be so small that CR are locked into traps
and carried away downstream through the shock.
The condition for CR escaping the trap before being overtaken by the shock is that
$
q_{\rm trap}> {u}/{c}
$
since the CR passes between the mirror $\sim c/u$ times in the time $L_{\rm trap}/u$ 
taken for the shock to overtake the trap.
Equation (41) suggests that for realistic SNR-relevant magnetic field configurations and shock velocities, $q_{\rm trap}$ safely exceeds this lower limit.
Once again, a more sophisticated calculation is needed to resolve these issues.

\begin{figure}
\includegraphics[angle=0,width=6cm]{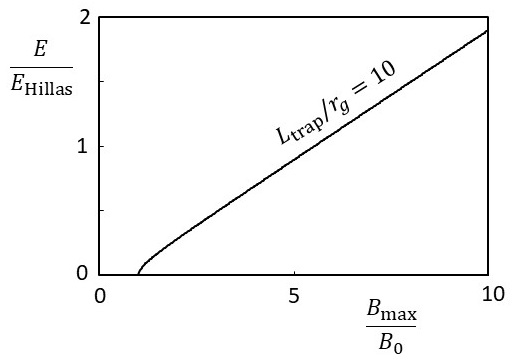}
\centering
\caption{
Ratio of the maximum CR energy to the Hillas energy $E_{\rm Hillas}=uBR$
as given  by equations (40) and (41) when $L_{\rm trap}/r_g=10$. 
}
\label{fig:figure5}
\end{figure}

\section{Mirror-dominated transport in other contexts}

This paper has been framed around the question of the maximum energy to which CR can be accelerated by shocks.
Our arguments may have a wider application to CR transport in the Galaxy and to UHECR transport between galaxies.

Reichherzer et al (2022; their figure 1) identify four different regimes of charged particle transport.
The mirror-dominated transport considered here corresponds most closely to their `mirror regime'.
Previous treatments of mirroring use a variety of approaches, variously starting from turbulence with a defined $k$-spectrum,
or `micro-mirrors', or calculating CR trajectories with a particle-in-cell code, or using a mixture of approaches 
(Cesarsky \& Kulsrud 1973, Lazarian \& Xu  2021, Barreto-Mota et al 2024, Zhang \& Xu 2024,  Reichherzer et al 2025).
How these compare with our use of VFP calculations and approximate configuration-space models
may be  a matter for future investigation.

Following from equations (39) and (41),
a diffusion coefficient for CR transport in the ISM might take the form 
$$
D_{\rm trap}=cL_{\rm trap}
\frac { B_{\rm min}}{B_{\rm max}}
 \left ( 1 - \frac {B_{\rm min}}{B_{\rm max}} \right )^ {-1/2} 
\eqno{(42)}
$$
where $B_{\max}$ and $B_{\min}$ are the characteristic maximum and minimum magnetic field
in any length of magnetic field line equal to the CR mean free path for angular scattering.
The justification for this dependence is that CR lose their memory of the magnetic field structure over distances larger than the CR mean free path.
Transport is controlled by the ratio $B_{\rm min}/B_{\rm max}$ in this distance.

Depending upon the $k$-spectrum of magnetic fluctuations in the ISM,
there may be some scalelength $L_{\rm trap}$ which dominates transport at each CR energy.
In this model, $B_{\rm max}/B_{\rm min}$ and $L_{\rm trap}$, both of which will in general be energy-dependent,  determine the CR diffusion coefficient and propagation time, which in turn
affects the composition and energy spectrum of CR arriving at the Earth.

Magnetic fields as large as 30-60nG have been discovered in filaments connecting galaxies (Vernstrom et al 2021).
A proton with an energy of 50EeV has a Larmor radius of $\sim 1 $Mparsec in such a field,
and a proton at the ankle in the UHECR spectrum at 5EeV has a Larmor radius of $\sim 100$kparsec.
Such a  spatially structured intergalactic field is sufficient to affect UHECR transport, playing a role
in forming the energy spectrum and generating anisotropy
(Condorelli et al 2023, Abdul Halim et al 2024, Marafico et al 2024).

UHECR anisotropy is structured on both large angular scales,
 such as the observed dipole,
and on  smaller angular scales (hotspots) as in the case of arrivals from the direction of Centaurus A 
or from echoes by magnetised galactic haloes (Bell \& Matthews 2022, Taylor et al 2023).
Transport dominated by localised mirrors may explain the co-existence of isotropy and  anisotropy on many scales.
It can produce large-angle scattering for UHECR encountering a mirror 
while at the same time allowing other UHECR to follow essentially straight line trajectories if they
do not encounter a mirror.
Mirror-dominated transport allows high-order multipole anisotropies to co-exist with a strong isotropic component and with low order dipole and quadrupole anisotropies. 
A conventional diffusion model with a uniform small-angle scattering rate is much more constraining in the anisotropy
mode-spectrum that it allows.
Future observations of  UHECR anisotropy may be able to distinguish between diffusion and mirror models of transport.

\section{Conclusions}
The theory of 
DSA is usually developed on the assumption that CR trajectories  are
determined by small-angle scattering by fluctuations in the magnetic field on scales smaller than the  CR Larmor radius.
With this assumption it can be shown that the maximum energy to which CR can be accelerated is the Hillas energy, $E=uBR$.
Except in the case of perpendicular shocks, the Hillas energy is only achieved if the CR mean free path is of the order of
the Larmor radius resulting in Bohm diffusion with a coefficient $D_{\rm Bohm}\sim r_g c$.
A non-resonant instability, driven by CR currents, amplifies the magnetic field with a wavelength that grows during amplification,
and saturates when the wavelength reaches the Larmor radius.
The combination of non-resonant field amplification and DSA satisfactorily accounts for CR acceleration by SNR to energies of the order of
a few hundred TeV, but it cannot account for acceleration to the knee at a few PeV.

Special sites such as very young SNR (none are known in the Galaxy at present), star-forming regions where the magnetic field is large and there are repeated supernovae and strong stellar winds, or the Galactic centre where similar conditions apply,
may account for PeV acceleration.
However, we instead question the assumption that small-angle scattering dominates CR transport.
We show that mirroring by magnetic fluctuations on scales larger than the Larmor radius can be equally as good 
as Bohm diffusion at confining CR close to a shock.
We show that the Hillas energy may be achievable without the generation of magnetic fluctuations on the Larmor scale.
Mirror-inducing magnetic fluctuations may be expected in the environment of SNR such as Cas A where the medium
upstream of the shock is disturbed by clumpy pre-supernova ejections and by fast knots.

MDSA is not as reliably efficient as standard DSA because the required mirrors may not always be 
present.  
Spasmodic acceleration may be expected, leading to a steeping of the spectrum beyond energies at which standard DSA operates.
The spectrum of gamma-rays produced by high energy CR will further steepen because CR preferentially
escape upstream of the MDSA process, thereby not entering the dense interior of the SNR where gamma-rays are
efficiently produced.
Hence, acceleration to PeV, and possibly higher energies, is not necessarily precluded by 
the observed steepening of the gamma-ray and CR spectrum of young SNR.

We also briefly consider the role of mirror-dominated transport in the ISM and the IGM, and suggest the possibility that 
the strong CR pressure gradients generated by mirrors may act to amplify the mirrors.


\section*{Acknowledgments}
JHM acknowledges a Royal Society University Research Fellowship.
AT acknowledges support from DESY (Zeuthen, Germany), a member of the Helmholtz Association HGF.
Computing resources were provided by STFC Scientific Computing Department’s SCARF cluster.
\vskip 0.2 cm


\section*{Data Availability}
The data underlying this article are available from the authors on reasonable request. 


\section*{References}
Abdul Halim A. for the Pierre Auger Consortium 2024, JCAP07(2024)094
\newline
Abeysekara A.U. et al, 2020, ApJ 894, 51
\newline
Abramowski A. for the HESS Collaboration, 2016, Nature 531, 476
\newline
Adams C.B. for the VERITAS Collaboration, 2021, ApJ 913, 115
\newline
Aharonian F. for the HESS Collaboration, 2024, ApJ 970, L21
\newline
Ahnen M.M. for the MAGIC Collaboration,  2017, MNRAS, 472, 2956
\newline
Arias M. et al,  2018,  Astron Astrophys 612, A110
\newline
Axford W.I., Leer E., Skadron G., 1977, Proc 15th Int Cosmic Ray Conf., 11, 132
\newline
Barreto-Mota L., de Gouveia Dal Pino E.M., Xu S., Lazarian A., 2024,
doi: 10.48550/arXiv.2405.12146
\newline
Beck R., 2015, Astron Astrophys Rev 24:4
\newline
Bell A.R., 1978, MNRAS 182, 147 
\newline
Bell A.R., 2004 MNRAS  353, 550
\newline
Bell A.R., Matthews J.H., 2022, MNRAS 511, 448
\newline
Bell A. R., Robinson A. P. L., Sherlock M., Kingham R. J., Rozmus W.,
2006, Plasma Phys Controlled Fusion, 48, R37
\newline
Bell A.R., Schure K.M., Reville B., Giacinti G., 2013, MNRAS 431, 415
\newline
Blandford R.D., Ostriker J.P., 1978, ApJ 221, L29
\newline
Cao Z. for the LHAASO XCollaboration, 2024, ApJ 961, L43
\newline
Cesarsky CJ., Kulsrud R.M., 1973,  ApJ 185, 153
\newline
Condorelli  A., Biteau J., Adam R., 2023, ApJ 957, 80
\newline
Cristofari P., 2021, Universe 7, 324
\newline
Decker R.B., 1993, JGR(A1) 98, 33
\newline
Duffy P., Kirk J.G., Gallant Y.A., Dendy R.O., 1995, Astron Astrophys 302, L21
\newline
Hillas A.M., 1984, ARA\&A 22, 425
\newline
Jokipii J,R., 1966, ApJ 143, 961
\newline
Jokipii J.R.,1982, ApJ 255, 716
\newline
Jokipii J.R., 1987, ApJ 313, 842
\newline
Krymsky G.F., 1977, Sov Phys Dokl, 22, 327
\newline
Lagage P.O., Cesarsky C.J., 1983a, A\&A 118, 223
\newline
Lagage P.O., Cesarsky C.J., 1983b, ApJ 125, 249
\newline
Lazarian A., Xu S., 2021, ApJ 923, 53
\newline
Lazarian A., Yan H., 2014, ApJ 784, 38
\newline
Lemoine M., 2022, Phys Rev Lett 129, 215101
\newline
Lemoine M., 2023, J Plasma Phys 89, 175890501
\newline
Marafico S., Biteau J., Condorelli A., Deligny O., Bregeon J., 2024, ApJ 972, 4
\newline
Milisavljevic D. et al, 2024, ApJ 965, L27
\newline
Muena C., Riquelme M., Reisenegger A., Sandoval A., 2024, Astron Astrophys 689, A216
\newline
Oka M., Makishima K., Terasawa T., 2025, ApJ 979, 161
\newline
Plotko P., van Vliet A., Rodrigues X., Winter W., ApJ 953, 129
\newline
Reichherzer P., Merton L., D\"{o}rner J., Becker Tjus J., Pueschel M.J., Zweibel E.G., 2022, SNAS 4:15
\newline
Reichherzer P., Bott A.,  Ewart R.J., Gregori G., Kempski P.,  Kunz M.W., Schekochihin A.A., 2025, 
Nat Astron, 
https://doi.org/10.1038/s41550-024-02442-1
\newline
Reville B., Bell, A.R., 2013, MNRAS 430, 2873
\newline
Taylor A.M., Matthews J.H.,  Bell A.R., 2023, MNRAS 524, 631
\newline
Tsunesada Y. for the Pierre Auger and Telescope Array Collaboration, 2021,
37th International Cosmic Ray Conference. 12-23 July 2021. Berlin, Germany, published 2022 at https://pos.sissa.it/cgi-bin/reader/conf.cgi?confid=395, id.337,
doi: 10.22323/1.395.0337 
\newline
Vernstrom T., Heald G.,  Vazza F., Galvin T.J., West J.L., Locatelli N., Fornengo N., Pinett E., 2021, MNRAS 505, 4178
\newline
Vieu T., Reville B., 2023, MNRAS 519, 136
\newline
Vieu T., Reville B., Aharonian F., 2022, MNRAS 515, 2256
\newline
Vink J., 2024, doi: 10.48550/arXiv.2406.03555 
\newline
Vink J., Laming J.M., 2003, ApJ 584, 758
\newline 
Vink J., Patnaude D.J., Casro D., 2022, ApJ 929, 57
\newline
V\"{o}lk H.J., Berezhko E.G., Ksenofontov L.T., 2005, Astron Astrophys 433, 229
\newline
Zhang C., Xu S, 2024, ApJ 975, 65
\newline
Zirakashvili V.N., Ptuskin V.S., 2008,  ApJ 678, 939
\newline

\end{document}